\documentclass[titlepage,12pt]{utarticle}
\usepackage{hyperref} 
\usepackage{amsmath,amsfonts}
\usepackage{array}
\long\def\omit#1{}

\numberwithin{equation}{section}

\newcommand{\homoquot}[3]{#3\Bigl\backslash\frac{#1}{#2}}
\newcommand{\smallhomoquot}[3]{#3\bigl\backslash\frac{#1}{#2}}

\begin{document}

\preprint{
UTTG--08--97\\
{\tt hep-th/9704087}\\
}

\title{On the Moduli Spaces of M(atrix)-Theory Compactifications}

\author{David Berenstein, Richard Corrado, and Jacques Distler
        \thanks{Research supported in part by the Robert A.\ Welch
        Foundation and NSF Grant PHY~9511632.}}
\oneaddress{ Theory Group, Department of Physics\\
        University of Texas at Austin\\
        Austin TX 78712 USA  \\ {~}\\
        \email{david@zippy.ph.utexas.edu}
        \email{rcorrado@zippy.ph.utexas.edu}
        \email{distler@golem.ph.utexas.edu}
        }

\date{April 10, 1997}

\Abstract{By identifying the moduli space of coupling constants in the
SYM description of toroidal compactifications of
M(atrix)-Theory, we construct the M(atrix) description of the moduli
spaces of Type IIA string theory compactified on $T^n$. Addition of
theta terms to the M(atrix) SYM produces the shift symmetries
necessary to recover the correct 
global structure of the moduli spaces. Up to $n=3$, the corresponding BPS
charges transform under the proper representations of the U-duality groups.
For $n=4,5$, if we make the \textit{ans\"atz} of including the BPS charges
corresponding to the wrapped M-theory 5-brane, the correspondence with Type
IIA continues to hold. However, for $n=6$, we find additional charges for
which there are no obvious candidates in M(atrix)-Theory. }

\maketitle
\renewcommand{\baselinestretch}{1.25} \normalsize

\section{Introduction}

One of the principles which drove the recent ``revolution'' in string
theory was Witten's discovery of M-Theory as a limit of Type IIA strings.
Building on results of Townsend~\cite{Townsend:revisited} and others, he
realized that the spectrum of BPS-saturated threshold bound states of
RR~charges in Type~IIA string theory were in direct correspondence with the
Kaluza-Klein spectrum of the $D=11$~supergravity multiplet with the
11th dimension compactified on a circle of radius~$r=\lambda^{2/3}$,
where $\lambda$ is the IIA~string
coupling~\cite{Witten:variousdims}. The eleven-dimensional theory
describing the Type~IIA theory at strong coupling later came to be
called M-Theory. Another crucial development was Polchinski's
identification of the RR~charged states with
D-branes~\cite{Polchinski:DBraneRR}, which provided a conformal field
theoretic description of strings in the presence of RR p-branes
(see~\cite{{Polchinski:NotesDbranes},{Polchinski:TASIDbrane}} for
discussions and more recent progress).

Witten soon showed that, for~$N$ coincident D-branes, the $U(1)^N$
RR~gauge group is enhanced to $U(N)$. The extra gauge bosons arise as the
lowest modes of
the open strings which stretch between the branes~\cite{Witten:BoundStrings}.
The low energy effective action one obtains is given by the dimensional
reduction of
${\cal N}=1$,
$D=9+1$ SYM to the D-brane world-volume. In this picture, separating
coincident D-branes is equivalent to breaking part of the $U(N)$~via the
Higgs mechanism in the SYM theory. At very strong coupling, the D-brane
states are lighter than any massive excitation of the string, and their
dynamics decouples at low energies. As shown by Danielsson, Ferretti, and
Sundborg~\cite{Danielsson:Dparticle} and Kabat and
Pouliot~\cite{Kabat:Zerobrane}, the truncation of the spectrum to the
lowest lying states reproduces the correct gravitational interactions
of these branes, crucial to their interpretation as Kaluza-Klein
supergravitons.

M(atrix)-Theory took shape when Banks, Fischler, Shenker, and
Susskind~\cite{BFSS:Conjecture} realized that all massive string
excitations as well as antibranes (with negative~$p_{11}$) decouple in
the infinite momentum frame in the 11-direction. The D0-branes are
interpreted as partons and the dynamics of~$N$ partons is exactly
described by $U(N)$~supersymmetric quantum mechanics. To recover
eleven-dimensional physics in the infinite momentum frame, one takes
the limit $N,r \rightarrow\infty$, $N/r\rightarrow\infty$, where~$r$ denotes
the radius of the 11-dimension.  M(atrix)-Theory, if it is correct,
provides the first non-perturbative formulation of M-Theory.

Toroidal compactifications of M(atrix)-Theory were investigated by
Taylor~\cite{Taylor:Compact}, where the effective action for
compactification on $T^d$ was shown to be $d+1$-dimensional SYM theory
with the dual torus $\widetilde T^d$ as its base space.  Several aspects of
wrapped membranes and T and U-duality in these M(atrix)-Theory
compactifications are discussed
in~\cite{{Susskind:Tduality},{Ganor:BranesFluxes},%
{Sethi:RotInv},{Rozali:Uduality},{Gopakumar:BPSMatStrings}}.
The states corresponding to different wrapped membranes can be
interpreted in terms of the topological quantum numbers of
time-independent classical solutions of the equations of
motion. In particular, the first Chern class (magnetic flux)
corresponds to wrapped D2-branes,  the second Chern class (instanton
number) corresponds to wrapped longitudinal
5-branes~\cite{Berkooz-Douglas:FiveBranes} (D4-branes in the IIA
theory~\cite{{Ganor:BranesFluxes},{Guralnik:Torons}}), and the third
Chern class was conjectured~\cite{Berenstein:MatrixVarious} to be the
correct description for the wrapped D6-brane~\cite{Banks:BranesMat} of
the IIA~theory.

An explicit construction of the moduli space of scalars for
M(atrix)-Theory compactifications is of great importance in evaluating
its validity and in making further progress in understanding the
larger role of M-Theory.  For toroidal compactifications beyond the
3-torus, the M(atrix) SYM theory is
non-renormalizable~\cite{{Berenstein:MatrixVarious},{Rozali:Uduality},%
{Banks:StringsMat},{Fischler:Shrinking}},
at least by power-counting.  In this paper, we focus on
properties which are independent of the particular dynamics of the
theories in question, namely the moduli space and BPS spectrum. We
provide a construction of the moduli spaces of M(atrix)-Theory
compactified on tori. For spacetime dimensions $D=7,\dots, 10$, we  have agreement  with the moduli
spaces and central charges of the type IIA string theory. For the simplest cases,  we give the explicit map between
the M(atrix) and SYM moduli. For $D=6$ and below, the central charge corresponding to the wrapped transverse 5-brane is missing from the M(atrix)-Theory description. If we make the \textit{ans\"atz} of including the 5-brane charges by hand, the central charges do assemble into the correct representations of the U-duality group. For $D=5$ and below, there are moduli which are missing from the M(atrix) description. Finally, in $D=4$, there are new central charges in IIA string theory and in M-Theory for which there no plausible candidates in M(atrix)-Theory.

\section{Toroidal Compactifications of Type~IIA String Theory}
\label{sec:string}

As a first step in constructing the moduli space of M(atrix)-Theory
compactifications on general manifolds, it is instructive to reproduce
the moduli space of toroidal compactifications of weakly coupled
Type~IIA string theory and review the standard
arguments for U-duality on the central charge of
these moduli.

{}From the conformal field theory point of view, the most general
toroidal compactification of the Type~IIA theory to $10-d$~dimensions
is given by an even self-dual Lorentzian lattice $\Lambda^{d,d}$,
which corresponds to left-movers and right-movers living on different
tori~\cite{{Narain:toroidal}}. After a Poisson
resummation, this is seen to be equivalent to giving the metric, $g_{\mu\nu}$,
and the antisymmetric tensor field,
$B_{\mu\nu}$, constant background values on the torus
\cite{{Narain-Sarmadi-Witten}}.

For simplicity, consider $T^2$. It is instructive to think of the
compactification as occurring in two stages. In the
first stage, compactification of
$X_1$ on a circle,  we generate two extra $U(1)$ gauge fields  which
correspond to the Kaluza-Klein reduction of the metric
\begin{equation}
g_{\mu\nu}\to g_{\mu\nu}\oplus g_{\mu 1}\oplus g_{11}
\end{equation}
and the antisymmetric tensor field
\begin{equation}
B_{\mu\nu}\to B_{\mu\nu}\oplus B_{\mu 1}.
\end{equation}

The parameter $g_{11}$ controls  the size of the circle. On the other
hand, $g_{\mu 1}$ and $B_{\mu 1}$ are $U(1)$ gauge fields in the
9-dimensional theory. On further compactification to eight dimensions,
$g_{22}$ will control the size of the second circle. The third component of
the metric on $T^2$,
$g_{12}$, can be viewed as a Wilson line for the corresponding $U(1)$
field. In a similar fashion, $B_{12}$ can be viewed as a Wilson line for the
$U(1)$ gauge field arising from $B_{\mu\nu}$. The net effect is that
string states which are charged under both $U(1)$s in the 1~or
2~dimensions correspond to the left and right moving modes of the
string in these directions. They are therefore described in terms of
their momentum and winding around the torus. The asymmetry between
left-movers and right-movers is evident because the Wilson lines
generate shifts in the quantization of momenta between the 1~and
2~directions which are proportional to the windings.

So far, the values of $A_{\mu}$ and $A_{\mu\nu\rho}$, which come from
the Ramond-Ramond sector and also give $U(1)$ gauge fields in lower
dimensions, have been totally ignored.  The elementary string doesn't
carry RR~charge, so the description of the moduli space is greatly
simplified.  This is somewhat fortunate, since it is not
possible to describe general RR backgrounds in terms of a worldsheet
SCFT. However, as one
takes the theory to strong coupling, RR~charged solitons of the string become
light, so that the effect of the RR~gauge fields can no longer be ignored. At
weak string coupling, these RR~charged states correspond to
D-branes~\cite{Polchinski:DBraneRR}, which do admit a conformal field
theoretic description.  When one
compactifies enough  dimensions, D$p$-branes can wrap around $p$-cycles and
give new point-like excitations in the low energy spectrum. Wilson lines for
these RR~gauge potentials generate additional shifts in the
quantization of momenta along compact directions. Moreover, invariance
under large gauge transformations makes the configuration space of
these gauge potentials compact. Under U-duality, all these states mix,
therefore the full symmetry of the string theory vacua is enlarged
with respect to the Narain compactifications.

The RR moduli on $T^n$ include $n$ Wilson lines of the RR
1-form and $n(n-1)(n-2)/3!$ periods of the RR 3-form.  We can
therefore express the dimension of the moduli space for
compactification on $T^n$, for $n\leq 4$, as
\begin{equation} \label{eq:IIAmoduli}
\begin{split}
\dim {\cal M}_{\text{IIA}[T^n]} = & ~1~\text{dilaton VEV}
+ n^2~\text{Narain moduli} + n~\text{Wilson lines of the RR 1-form} \\
& + \frac{n(n-1)(n-2)}{3!}~\text{periods of the RR 3-form}.
\end{split}
\end{equation}
For larger $n$, there are additional scalars obtained by dualizing the
1 and 3-forms. For example, in five dimensions $A_{\mu\nu\lambda}$ dualizes
into a scalar. Similarly, for $n\geq 7$, there are moduli generated by
dualizing the 1-form $A_\mu$ into a scalar.  Therefore, when $n>5$ there
are, in addition to the moduli counted in~\eqref{eq:IIAmoduli},
\begin{equation} \label{eq:addIIAmoduli}
\begin{split}
& \frac{n(n-1)(n-2)(n-3)(n-4)}{5!}~\text{``duals'' of the RR
3-form} \\
& + \frac{n(n-1)(n-2)(n-3)(n-4)(n-5)(n-6)}{7!}~\text{``duals'' of the
RR 1-form} .
\end{split}
\end{equation}

The local structure of the moduli space is dictated by low-energy
supergravity considerations to be one of the homogeneous spaces listed in
Table~\ref{tab:disint}. The true moduli space is obtained by modding out by
the U-duality group~\cite{Hull:Unity}.
These U-dualities act on the central charge of the theory.
Let us recall how U-duality is realized in the string and M-Theory pictures,
following~\cite{{Hull:Unity},{Witten:variousdims}}. We will explicitly
consider the cases of eight and four spacetime dimensions, then recall the
group disintegration properties~\cite{Julia:Disintegrations} to summarize the results in intermediate dimensions.
\begin{table}[bht]
\begin{center}
\begin{picture}(150,220)
\put(-165,100){ \setlength{\extrarowheight}{12pt}
\begin{tabular}{|c|c|c|c|}
\hline
$D$ & Moduli Space & U-duality Group, $\Gamma$  & Rep.\ of the
Central Charge under $\Gamma$ \\
\hline
4 &$\smallhomoquot{E_{7(7)}}{SU(8)}{\Gamma}$& $E_{7(7)}(\BZ)$ &
$\mathbf{56}$  \\
5 &$\smallhomoquot{E_{6(6)}}{Sp(4)}{\Gamma}$& $E_{6(6)}(\BZ)$ &  1+
$\mathbf{27} $
\hspace*{5mm}  1+ $ \widetilde{27}$  \\
6 &$\smallhomoquot{SO(5,5)}{SO(5)\times SO(5)}{\Gamma}$& $SO(5,5,\BZ)$
& $1+10+\mathbf{16}$
\hspace*{5mm} $1+10+\widetilde{16}$  \\
7 &$\smallhomoquot{SL(5)}{SO(5)}{\Gamma}$& $SL(5,\BZ)$  &
$1+5+\mathbf{10}$ \hspace*{5mm} $1+\widetilde{5}+10$ \\
8
&$\Gamma\bigl\backslash\bigl(\frac{SL(3)}{SO(3)}\times
\frac{SL(2)}{SO(2)}\bigr)$&
$SL(3,\BZ)\times SL(2,\BZ)$  &
$(1,1)+ (\widetilde{3},1)+ (\mathbf{3},\mathbf{2})$ \hspace*{1mm}
$(1,1)+ (3,1)+ (\widetilde{3},2)$  \\
9 &$\smallhomoquot{GL(2)}{SO(2)}{\Gamma}$& $SL(2,\BZ)\times \BZ_2$ &
$1+2+\mathbf{1}+\mathbf{2}$
\hspace*{2mm} $1+\widetilde{2} +1 + \widetilde{2}$  \\
\hline
\end{tabular}   }
\put(205,152){\vector(-1,-1){15}}
\put(205,152){\vector(3,-1){30}}
\put(187,122){\vector(0,-2){15}}
\put(187,90){\vector(0,-2){15}}
\put(187,60){\vector(0,-2){15}}
\put(187,27){\vector(0,-2){15}}
\end{picture}
\caption{U-duality and central charges. The central charge in
$D$ spacetime dimensions belongs to the bold representation of the U-duality
group and the arrows indicate how the representation decomposes under the
U-duality group in $D+1$ dimensions.} \label{tab:disint}
\end{center}
\end{table}

First we consider type IIA string theory on a 2-torus. Central charges
couple to the $U(1)$ vectors obtained by saturating all but one index
of a $p$-form. From the NS-NS sector, there are four such $U(1)$s,
two from $g_{\mu a}$ and another two from $B_{\mu a}$, collectively
forming the vector~$\mathbf{4}$ of the T-duality group,
$SO(2,2,\BZ)$. In the RR sector, there are two more $U(1)$s, $A_\mu$
and $A_{\mu 12}$, transforming as the spinor~$\mathbf{2}$ of
$SO(2,2,\BZ)$. Together these form the
$(\mathbf{3}, \mathbf{2})$ representation of the U-duality group
$SL(3,\BZ)\times SL(2,\BZ)$. In the dual picture of M-Theory on a
3-torus, the gauge fields in the $(\mathbf{3}, \mathbf{2})$ are the three
``electric" gauge fields which arise from the metric,
$g_{\mu a}$, and the 3 ``magnetic" gauge fields which arise from the 3-form,
$A_{\mu ab}$.

In four spacetime dimensions, the IIA theory has 12 $U(1)$s in the NS-NS
sector, with 24 corresponding electric and magnetic charges, while in
the RR sector there are 16~$U(1)$s, with 32 corresponding electric and
magnetic charges.  Collectively, the charges form the 
$\mathbf{56}$ of the U-duality group, $E_{7(7)}$. For M[$T^7$], there
are 7~$g_{\mu a}$ and 21~$A_{\mu ab}$, with the electric and magnetic
charges coupling to these again generating the $\mathbf{56}$ of
$E_{7(7)}$. The classification of the objects which carry these charges
in the IIA and M-Theory pictures is given in Table~\ref{tab:charges}.
\begin{table}[htb]
\begin{center}
\begin{tabular}{|c|c|c|}
\hline
IIA[$T^6$] & M[$T^7$] & M(atrix)[$\widetilde{T}^7$] \\
\hline
\begin{tabular}{c}
1 D0-brane charge \\ 6 momentum modes
\end{tabular} & 7 momentum modes & 7 electric fluxes \\
\hline
\begin{tabular}{c}
6 winding modes \\ 15 wrapped D2-branes
\end{tabular} & 21 wrapped 2-branes & 21 magnetic fluxes \\
\hline
\begin{tabular}{c}
6 wrapped NS-NS 5-branes \\ 15 wrapped D4-branes
\end{tabular} & 21 wrapped 5-branes & ? \\
\hline
\begin{tabular}{c}
1 wrapped D6-brane \\ 6 KK monopoles
\end{tabular} & 7 KK monopoles & ?? \\
\hline
\end{tabular}
\caption{The $U(1)$ charge-carrying pointlike states in four spacetime
dimensions in the IIA[$T^6$], M[$T^7$], and
M(atrix)[$\widetilde{T}^7$] pictures.} \label{tab:charges}
\end{center}
\end{table}

In higher spacetime dimensions, the structure of the U-duality groups
can be understood from the decomposition of the  $\mathbf{56}$ of
$E_{7(7)}$. The sequence of disintegrations is given in
Table~\ref{tab:disint}.

\section{The Moduli Spaces of M(atrix)-Theory on Tori}

M(atrix)-Theory allegedly contains $D=11$ supergravity in its low
energy spectrum, and moreover describes all the physics relevant to
the strong coupling limit of Type~IIA string theory. Therefore, the
moduli space of string vacua should arise naturally in the language of
M(atrix)-Theory. In this section, we will explicitly
compute these moduli spaces to verify this claim.

In discussing M(atrix)-Theory, we consider the case in which the spatial
dimension defining the infinite momentum frame is non-compact,
{\it i.e.}, we consider the full $N,r_{\text{IMF}} \rightarrow\infty$,
$N/r_{\text{IMF}}\rightarrow\infty$ limit. Therefore, when comparing a
toroidal compactification with the IIA theory, we do not compactify
the infinite momentum frame and one of the radii of the torus will
correspond to the IIA dilaton. Finally, in light of the fact that the
SYM theory that describes compactified M(atrix)-Theory is formulated
on the dual torus, we employ a notation in which the duality
relationships are
\begin{equation}
\text{IIA}[T^{d-1}] \sim \text{M}[T^d] \sim
\text{M(atrix)}[\widetilde{T}^d].
\end{equation}

M(atrix)-Theory on a torus $\widetilde{T}^d$ is precisely SYM on
$\widetilde{T}^d\times \BR$~\cite{Taylor:Compact}. Part of the moduli
space is described by the moduli of constant metrics of unit volume on
the $d$-torus modulo diffeomorphisms. The space of such metrics is
$SL(d,\BZ) \backslash SL(d,\BR) / SO(d) $. Note -- and this is very
important -- that $SL(d,\BZ)$ is not a symmetry of the Yang-Mills
theory on $\widetilde T^d$. However, we shall see that $SL(d,\BZ)$ is
a symmetry of the spectrum of BPS states and, very likely, of their
interactions\footnote{To lowest order, the interactions between the BPS states arise from $F_{\mu\nu}^4$ terms. The resulting scattering amplitudes are proportional to  the square of  a bilinear in \textit{differences} of BPS charges. The bilinear is formed by dotting the differences of BPS charges into the same quadratic form as appears in the BPS mass formula. The resulting formula is naturally U-invariant.}.

In addition, there are two more \textit{dimensionful} parameters: the
size of the torus, $V=\text{Vol}(\widetilde{T}^d)$, and the coupling
constant, $g$, with dimension $[g]\sim (\text{mass})^{(3-d)/2}$. One
combination of these sets the mass scale for the spacetime theory. The
other combination, the dimensionless coupling constant
\begin{equation} \label{eq:dimcoup}
\tilde{g} = g V^{(d-3)/2d},
\end{equation}
is a modulus of the theory (see also the discussion
in~\cite{Fischler:Shrinking}).

In describing the moduli space, we must also give an exact description of
how the momentum labels are found for the compact directions, as well
as an explicit construction of the deformations of the field theory
that give rise to the shifts in momenta found in the~IIA theory. We
will now show explicitly how the quantization of momenta in the
compact directions is obtained.

First, notice that, as $H_1(\widetilde{T}^d,\BZ)=\BZ^d$, one can have
Wilson lines for the $U(1)$ gauge fields along each of these
cycles. These correspond to the values of the zero-modes of the gauge
fields. As the SYM theory is gauge invariant, the zero-modes,
$A^0_\mu$, satisfy
$A^0_\mu dx^\mu \in H^1(\widetilde{T}^d,\BR)$. However, since the
zero-modes are only well defined modulo the shifts generated by large
gauge transformations, we really should refine this statement to
\begin{equation}
A^0_\mu dx^\mu \in H_1(\widetilde{T}^d,\BR)/\BZ^d.
\end{equation}
Hence the gauge field configuration space of Wilson lines is a torus
that has the same shape as~$T^d$. Conjugate momenta will be
proportional to $\dot{A}_\mu dx^\mu = F_{0\mu}$, where we have taken
$A_0=0$, and are quantized in units of the dual
torus~$\widetilde{T}^d$. Therefore the electric fluxes provide the
correct description of the momenta in the compact
directions.

In IIA compactifications, Wilson lines produce shifts in momenta which
are proportional to the winding numbers of the string.  Here in the
M(atrix) SYM theory, the electric flux is also allowed shifts in
quantization. These shifts are generated by the addition of a total
time derivative that does not affect the equations of motion, but that
does affect the quantization of momenta.  This will be our approach
for M(atrix)-Theory. We will add total time derivatives to the action
and we will interpret each one of these terms as an allowable
deformation. As we will see later, counting all these terms properly
will give the coordinates of the IIA~moduli space. They will also have
all of the shift symmetries of the corresponding moduli, which in the
string theory correspond to large gauge transformations that generate
shifts of integral multiples of $2\pi i$ in the world-volume action of
membranes wrapped around non-trivial cycles.

\subsection{M(atrix)-Theory in Ten Dimensions}
One phenomenon that we will meet as we consider SYM theory in various
dimensions is the possibility of adding topological terms to the
action, that is, the integrals of various characteristic classes, $\CP(V)$, of
the vector bundle $V$. Not all of these will lead to sensible physics. We need
to require that widely separated clusters of D0-branes should approximately
decouple. In the M(atrix)-Theory language, this means that when the vector
bundle $V$ is a direct sum, the action should factorize. For the topological
terms, then, we need to restrict ourselves to characteristic classes which
satisfy
\begin{equation}\label{eq:charclassone}
\CP(V_1\oplus V_2)=\CP(V_1)+\CP(V_2)\, .
\end{equation}
Furthermore, we wish to respect the charge-conjugation
symmetry\footnote{In M-Theory, this is $CP$, where $C: A\to -A$ and $P:x^i\to -x^i$, for $i=1,\dots 9$.} which, in the M(atrix) theory language, exchanges the
vector bundle $V$ with its dual,
\begin{equation}\label{eq:charclasstwo}
\CP(V)=\CP(V^*)\, .
\end{equation}
For $U(N)$ bundles, the first restriction~\eqref{eq:charclassone} means that we
should restrict ourselves to considering the Chern character, $ch(V)$.  The
second restriction~\eqref{eq:charclasstwo} means that we should consider only
the even\footnote{The first Chern character,
$ch_1(V)=c_1(V)=\frac{1}{2\pi}\int \text{Tr}\, F$, actually has a direct
connection to the Galilean invariance of the theory. The theta angles
for this term are, in fact, the angles between
the sides of the torus and the light-cone direction. Since this term
only couples to the $U(1)$ component of the $U(N)$ gauge group, this
term is decoupled from the interactions, therefore providing a crucial
test of the Galilean invariance.} Chern characters,
\begin{equation*}
ch_2(V)= \frac{1}{8\pi^2}\text{Tr}\, F\wedge F, \quad
ch_4(V)=\frac{1}{4!(2\pi)^4}\text{Tr}\, F\wedge F\wedge F\wedge F,
\end{equation*}
\textit{etc.} These terms will only start to make their appearance
once we have compactified a sufficient number of dimensions.

The counting of moduli for the case of M(atrix)-Theory on $S^1$ is
rather trivial, since there is only one modulus, namely the SYM
coupling constant. This corresponds to the VEV of the dilaton in the
Type IIA string theory and we can use~\eqref{eq:dimcoup} to obtain this
correspondence. We compare the value of the D0-brane tension, 
\begin{equation} \label{eq:dzerotension}
T_{D0}= M_s e^{-\phi^{(10)}},
\end{equation}
where $\phi^{(10)}$ is the 10-dimensional dilaton, with the M(atrix)
value $T_{D0} = 1/r$. Noting that the eleven-dimensional Planck mass
is related to the string scale via $M_P = e^{-2 \phi^{(10)}/9} M_s$, we
find the expected relationship~\cite{Witten:variousdims}
\begin{equation} \label{eq:releven}
r = M_P^{-1} e^{2\phi^{(10)}/3}.
\end{equation}

\subsection{M(atrix)-Theory in Nine Dimensions}

Now consider M(atrix)[$\widetilde{T}^2$], whose moduli are described
by the dimensionless coupling constant and the 2 parameters of a unit
volume metric on the 2-torus. Equivalently, instead of the volume-one
metric, we can consider the complex structure of the torus, which
encodes the same information. Therefore, the dimension of the
moduli space is equal to that for IIA[$S^1$] as obtained
{}from~\eqref{eq:IIAmoduli}. In fact, the global structure is the same in both
cases. The complex structure, $\tau_{\text{cplx.}}$, lies in the fundamental
domain
${\cal F}$ in the upper half plane, while the coupling constant is a positive
real number, so
\begin{equation}
{\cal M}_{\text{M(atrix)}[\widetilde{T}^2]} = {\cal F}\times \BR^+ .
\end{equation}
We can arrange the moduli as
\begin{equation} \label{eq:ninedimmod}
\begin{array}{cl}
(w + i e^{-\phi}, M_s^{-1} R) &
\text{for IIA}[S^1] \\
(\widetilde{\varphi} + i e^{-\phi},   2M_s R^{-1}) &
\text{for IIB}[S^1] \\
(\tau_{\text{cplx.}}, 1/\tilde{g}^2) &
\text{for M(atrix)}[\widetilde{T}^2],
\end{array}
\end{equation}
where $w =\oint A\cdot dx$ is the RR Wilson line of the IIA theory,
$\widetilde{\varphi}$ is the RR scalar of the IIB theory, $\phi$ is the
9-dimensional dilaton, and $R$ is the radius of the circle in the IIA
compactification. In each case, the moduli space is
\begin{equation}
\homoquot{GL(2,\BR)}{SO(2)}{SL(2,\BZ)\times \BZ_2} =
{\cal F}\times \BR^+,
\end{equation}
so that the M(atrix)-Theory moduli space exactly agrees with that of
the string theories. The
equations~\eqref{eq:ninedimmod} give the mapping between
the parameterizations and we find that~\eqref{eq:releven} is
satisfied. Additionally, we see, heuristically, the correspondence
between the $\tilde{g}\rightarrow \infty$ limit of the SYM theory and
the Type~IIB theory in ten dimensions.

The $SL(2,\BZ)$ component of the U-duality group in this picture acts
on the two electric fluxes and one magnetic flux, which form the
$\mathbf{2}+\mathbf{1}$ representation.  This is not a symmetry of the
2+1-dimensional SYM action, but is rather a symmetry of the BPS spectrum.

\subsection{M(atrix)-Theory in Eight Dimensions}

M(atrix)[$\widetilde{T}^3$] has 6~moduli corresponding to the coupling
constant and the metric on the torus $\widetilde{T}^3$. The question
now is, what corresponds to the VEV of $B_{12}$ in M(atrix)-Theory?
The effect of this VEV in string theory is to shift the quantization
of momenta in the partition function of wrapped strings along the
2-cycle dual to $B$. In M-Theory, these strings correspond to 2-branes
which are wrapped around the whole of $T^3$. In M(atrix)-Theory these
wrapped membranes are represented by states that carry magnetic flux
in, say, the 1,2-direction with momentum along the 3-direction. In the
string picture, this corresponds to winding in the 2-direction with
momentum along the 3-direction. Exchange of winding and momentum is
consistent with a permutation of the~2 and 3-directions in
M(atrix)-Theory. Note that, if momentum is carried along the
1-direction, from the IIA~point of view, this state is a soliton, so
it corresponds to a wrapped D2-brane.

For these wrapped string states, the shift in momentum requires states
that are both wrapped and which carry momentum. As a result, in the
partition function at fixed momentum and winding, we will have an
action proportional to
\begin{equation}
v^2 + w^2 + v\cdot b\cdot w,
\end{equation}
where $v$ corresponds to the classical velocity around the cycle, $w$
is the winding, and $b$~is the expectation value of
$B_{\mu\nu}$. Since $w$ is held fixed and this is a total time
derivative, we obtain the expected shift in the quantization of the momenta
\begin{equation}\label{eq:pshift}
\Delta p\sim b\cdot w.
\end{equation}

Finally, as we are in 3+1 dimensions, we have our first opportunity to add a
topological term to the action. We add
\begin{equation} \label{eq:fsquared}
2\pi b \int ch_2(V)= \frac{b}{4\pi} \int \text{Tr} \, F\wedge F\, .
\end{equation}
As $2\pi b$ is an angle, $b\sim b+1$, and combines
with the Yang Mills coupling to form the complex gauge coupling
$b+\frac{4\pi i}{\tilde{g}^2}$ of the $\CN=4$, $D=4$ SYM.
The shift~\eqref{eq:pshift} in the momentum is, in the context of M(atrix)-Theory, simply a manifestation of the Witten effect~\cite{Witten:ThetaVacua}, whereby the electric charge of a state is shifted by a term proportional to $\theta$ times the magnetic charge.

Now the T-duality of M(atrix)[$\widetilde{T}^3$] was studied
in~\cite{{Susskind:Tduality},{Ganor:BranesFluxes}}, where it was
noted that, since the SYM is conformal, the $SL(2,\BZ)$ part of the
U-duality group is insured by S-duality\footnote{It is of
incredibly good fortune that M(atrix)-Theory is based on $U(N)$, a
self-dual group. For a group which is not self-dual, only a subgroup
of $SL(2,\BZ)$ preserves the gauge group under
S-duality~\cite{Vafa:SDuality}.} . In this case, the $SL(2,\BZ)$ S-duality acts
on the complex
coupling constant $b+\frac{4\pi i}{\tilde{g}^2}$ and there is a
$SL(3,\BZ)$ action on the 3 electric fluxes and 3 magnetic fluxes,
which form the $(\mathbf{3}, \mathbf{2})$ representation of the
combined U-duality group $SL(3,\BZ)\times SL(2,\BZ)$. Even though the
SYM theory is conformal, the $SL(3,\BZ)$ component of the U-duality
group is not a symmetry of the action. It is only a symmetry of the
BPS spectrum. 

We would like to see the explicit correspondence between the SYM
moduli and those of the IIA picture. The 7~IIA[$T^2$] moduli are, from
our discussion in section~\ref{sec:string}, given by the
dilaton, $\phi$, four Narain moduli (the complex structure,
$\tau=\tau_1+i\tau_2$, and $\rho=B+ig$), as well as the two RR Wilson
lines, $w_a = \oint A_a dx^a$. The T-duality group is $SL(2,\BZ)\times
SL(2,\BZ)$, where the
first factor acts on $\tau$ and the second on $\rho$. This gets promoted to the
full U-duality group, $SL(3,\BZ)\times SL(2,\BZ)$. We would like to
see how the five additional moduli combine with $\tau$ to form the homogeneous
space  $SL(3,\BR)/SO(3)$.

To this end, we consider the action of the $SL(2,\BZ)\subset SL(3,\BZ)$
subgroup, $\Bigl(\begin{smallmatrix}a&b&0\\ c&d&0\\
0&0&1\end{smallmatrix}\Bigr)$, which acts as
$\tau\to\frac{a\tau+b}{c\tau+d}$ on the complex structure of the
torus\footnote{The $SL(2,\BZ)$ does not act faithfully on $\tau$, but it does
act faithfully on the Wilson lines, $w_{1,2}$.}.   It will prove useful to
employ the GKD decomposition of an
$SL(n,\BR)$ matrix,
$M$, into the product
\begin{equation}\label{eq:GKD}
M=U D {\cal O},
\end{equation}
where $U$ is upper-triangular (with 1s on the diagonal), $D$ is
diagonal with $\det D=1$, and ${\cal O}$ is orthogonal. This
decomposition is the natural one to use to parameterize the quotient
space $SL(3,\BR)/SO(3)$. To find the dependence on $\tau_1$ and the
$w_a$, we consider the left action of the Borel subgroup,
$\Bigl(\begin{smallmatrix}1&n & 0\\ 0 & 1&0\\ 0&0&1\end{smallmatrix}\Bigr)$ of
$SL(2,\BZ)$ on
$M$. This affects only the matrix $U$ in the above decomposition. Acting with
$S=\Bigl(\begin{smallmatrix}0&-1&0\\ 1&0&0\\ 0&0&1\end{smallmatrix}\Bigr)$, the
result can be brought back into the  form~\eqref{eq:GKD} by the right action
of an $SO(3)$ matrix. This determines the dependence on $\tau_2$. Finally,  the
dilaton
dependence is obtained by analyzing the mass spectrum obtained from the
quadratic form on the $(\mathbf{3},\mathbf{2})$ representation,
\begin{equation}
MM^T \otimes \frac{1}{\text{Im}\, \rho}
\begin{pmatrix}
|\rho|^2 & \rho_1 \\
\rho_1 & 1
\end{pmatrix}
\bigl( M_{pl}^{(8)}\bigl)^2 ,
\end{equation}
and comparing with the string
result~\eqref{eq:dzerotension}. We find that $SL(3,\BR)/SO(3)$ is
parameterized by
\begin{equation} \label{eq:GKDeight}
M=
\begin{pmatrix}
1 & \tau_1 & w_2 \\
0 & 1 & w_1 \\
0 & 0 & 1
\end{pmatrix}
\begin{pmatrix}
\sqrt{\tau_2} e^{\phi/3} & 0 & 0 \\
0 & \frac{1}{\sqrt{\tau_2}} e^{\phi/3}& 0 \\
0 & 0 & e^{-2\phi/3}
\end{pmatrix},
\end{equation}
where now $\phi$ is the eight-dimensional dilaton. From the IIA string
point of view, this is rather surprising, since the theory on the
2-torus actually possesses the symmetry of an eleven-dimensional
theory on the 3-torus.  A metric of unit volume on the 3-torus
is determined from  $g_{T^3} = M M^T = U D^2 U^T$. Therefore the
radius of the eleventh dimension, \eqref{eq:releven},
is manifest in this treatment. With hindsight, we, of course, realize
this as the manifestation of
M-Theory~\cite{{Townsend:revisited},{Witten:variousdims}}.

To summarize, the complex coupling constant of the SYM theory,
$b+\frac{4\pi i}{\tilde{g}^2}$, directly maps to the IIA modulus $\rho$. The
mapping between the $SL(3,\BR)/SO(3)$ M(atrix) SYM moduli, namely the
components of the metric on the 3-torus, and the corresponding IIA
moduli is obtained by comparing~\eqref{eq:GKDeight} with the metric
of unit volume on the 3-torus, $g = MM^T$, where
\begin{equation}
M =
\begin{pmatrix}
1 & \frac{ g_{12}g_{33} - g_{13}g_{23} }{ g_{22}g_{33}
- g_{23}^2 }
&  \frac{g_{13}}{g_{33}} \\
0 & 1 & \frac{g_{23}}{g_{33}} \\
0 & 0 & 1
\end{pmatrix}
\begin{pmatrix}
\frac{1}{\sqrt{g_{22}g_{33} - g_{23}^2} } & 0 & 0 \\
0 & \sqrt{ g_{22} - \frac{g_{23}^2}{g_{33} } } & 0 \\
0 & 0 & \sqrt{g_{33}}
\end{pmatrix} .
\end{equation}

\subsection{M(atrix)-Theory in Seven Dimensions}
\label{ssect:sevendim}

M(atrix)[$\widetilde{T}^4$] has nine moduli which describe the unit
volume metrics on the dual 4-torus in addition to the coupling
constant. The other four parameters (which in M-Theory correspond to
the different $A_{ijk}$ cycles) are given by integrals
\begin{equation} \label{eq:thetathree}
\sum_{e_{ijk}} \frac{A_{ijk}}{4\pi} \int dt \int_{e_{ijk}} \text{Tr}\, F\wedge
F,
\end{equation}
where $e_{ijk}$ is a basis for the homology 3-cycles of $T^4$. There
are four such 3-cycles. Again, the quantities $2\pi A_{ijk}$ are angles in
the same manner as the QCD $\theta$ angle. They multiply
topological invariants, and therefore don't modify the equations of
motion of M(atrix)-Theory, except at the expected shifts in the
momentum quantization.

Notice that, at this point, we are dealing with a
5-dimensional gauge field theory and the renormalizability of the
theory is doubtful. In four dimensions, this theory is
finite to all orders in perturbation theory, and the low energy
effective action receives no perturbative
corrections. In~\cite{Berenstein:MatrixVarious}, the 1-loop
contribution to the $\beta$-function was studied in these SYM
theories. In 5, 6, and 7~dimensions, one could explicitly show that
the theory does not get renormalized at 1-loop. However, in
8~dimensions a logarithmic divergence was found in $F_{\mu\nu}^4$ at
1-loop, so that the perturbation theory breaks down. This divergence
was related to the IR divergence present in gravity in four spacetime
dimensions due to massless particle exchange, but, in general, these
divergences should convince one to take perturbative calculations with
a grain of salt.

Beyond one loop, there is little explicitly known. However, some insight can
be gained by considering the heterotic string compactified on a torus. In
that theory\footnote{We thank Vadim Kaplunovsky for
discussions on this point.}, there are no perturbative corrections (finite or
infinite) to
$F_{\mu\nu}^2$. There are, of course, finite corrections to $F_{\mu\nu}^4$
and higher terms. If we view the string theory as a short-distance cutoff of
the (maximally-supersymmetric) SYM theory, we may expect some of
these corrections to diverge as we send the cutoff away. However, since there
were no finite corrections to $F_{\mu\nu}^2$, we expect it to stay
unrenormalized, even as we take the cutoff away. The next dangerous term is
$F_{\mu\nu}^4$, but simple power counting says that this is an irrelevant
operator for dimensions less than 8. One suspects that the string theory
result is a consequence of the SYM having maximal supersymmetry.
If so, one might expect that the result might hold for the $U(N)$,
$N\rightarrow\infty$, theories that we are actually interested in.

These observations suggest that the specific IR behavior of these
gauge theories must play a significant
role~\cite{{Rozali:Uduality},{Banks:StringsMat},{Fischler:Shrinking}}.
It would therefore be very interesting to have a detailed description
of the dynamics of these theories, such as that which would be
obtained by relating them to IR fixed points of other
theories. Rozali~\cite{Rozali:Uduality} has made the
extremely interesting observation that the $4+1$-dimensional
M(atrix)-Theory may be related to an IR fixed point of a theory in
$5+1$ dimensions~\cite{Seiberg:SixDim}. Further discussion along these
lines may be found in~\cite{{Banks:StringsMat},{Fischler:Shrinking}}.

Whatever the status of its renormalizability,  one still expects that M(atrix)
SYM contains a consistent description of the moduli and BPS states. Moreover,
all of this information is encoded in quantities that are well behaved in the
IR limit. We will therefore focus our discussion on the BPS properties, which
do not depend on the short-distance behaviour of the theory.

We would like to know what manifold the moduli of the SYM theory
parameterize. It is obvious that the components of the metric
parameterize an $SL(4,\BR)/SO(4)$ subspace, but the theta angles
should enhance this. We can see how this occurs by considering the
U-duality action on the central charges. Under the $SL(4,\BZ)$ group
of torus deformations, the 4~electric fluxes fall into the
vector~$\mathbf{4}$ and the 6~magnetic fluxes compose the
antisymmetric tensor~$\mathbf{6}_a$. This is the natural decomposition
of the antisymmetric tensor~$\mathbf{10}_a$ of $SL(5,\BZ)$ under its
$SL(4,\BZ)$ subgroup. Therefore these charges naturally
reproduce the correct representation of the U-duality group, as
indicated by Table~\ref{tab:disint}.

Now the four theta angles in~\eqref{eq:thetathree} form
a vector, $a_i=\epsilon_{ijkl} A^{jkl}/3!$  under $SL(4,\BZ)$
transformations. Therefore the theta angles appear in the $U$ factor
of the decomposition of $SL(5,\BR)/SO(5)$ and the $D$ factor is
determined up to a function of $\tilde{g}$,
\begin{equation}
M = U D =
\begin{pmatrix}
U_4 & a_i \\
0 & 1
\end{pmatrix}
\begin{pmatrix}
D_4 f(\tilde{g})^{-1} & 0 \\
0 & f(\tilde{g})^4
\end{pmatrix},
\end{equation}
where 
\begin{gather*}
U_4= \begin{pmatrix} 
1 & 
\frac{ (g_{12}g_{44} -  g_{14}g_{24})(g_{33} g_{44}-  g_{34}^2) -
\left( g_{13}g_{44} - g_{14} g_{34} \right)
\left( g_{23}g_{44} -  g_{24} g_{34} \right)
}{ 
(g_{22}g_{44}- g_{24}^2)(g_{33}g_{44} -  g_{34}^2) 
- (g_{23}g_{44} -  g_{24} g_{34})^2  } 
&  \frac{g_{13}g_{44} -  g_{14} g_{34}}{g_{33} g_{44}
- g_{34}^2 } 
& \frac{g_{14}}{g_{44}} \\
0 & 1 & \frac{g_{23}g_{44} - g_{24} g_{34} }{g_{33} g_{44}
- g_{34}^2 }
& \frac{g_{24}}{g_{44}}\\ 
0&0&1& \frac{g_{34}}{g_{44}} \\
0 & 0 & 0 & 1 
\end{pmatrix}  
\\
\begin{split}
D_4= \text{diag}\Biggl(& 
\tfrac{1}{\sqrt{ \left( g_{22}g_{44} - g_{24}^2 \right) 
\left( g_{33}g_{44} - g_{34}^2\right) 
- \left( g_{23}g_{44} - g_{24} g_{34} \right)^2 } },
\sqrt{\tfrac{(g_{22}g_{44} - g_{24}^2 )(g_{33}g_{44} 
- g_{34}^2)
- (g_{23}g_{44} -  g_{24} g_{34})^2  }{g_{33}g_{44} 
- g_{34}^2 } },\\ 
&
\sqrt{\tfrac{g_{33}g_{44} -  g_{34}^2 }{ g_{44} } } ,\sqrt{g_{44}} \Biggr)
\end{split}
\end{gather*}
correspond to the decomposition using~\eqref{eq:GKD} of the
metric on the 4-torus, $g_4 = U_4 D^2_4 U_4^T$. The coupling constant
dependence may be
determined by realizing that the quadratic form $MM^T$ extends to a
quadratic form on the antisymmetric tensor of $SL(5,\BR)$. This
yields the correct BPS mass formula for the charges. With
$g_{ij}$ a metric on the 4-torus of unit volume, we find, up to an
overall constant which fixes the scale of the Hamiltonian, 
\begin{equation}
{\cal M}^2 \sim
\tilde{g}^2 g_{ij} (n^i + A^{ikl} n_{kl})(n^j + A^{jrs} n_{rs})
+ \left( \frac{2\pi}{\tilde{g}} \right)^2 g^{ij} g^{kl} n_{ik} n_{jl} ,
\end{equation}
where the $n^i$ and $n_{ij}$ are integers, so that
$f(\tilde{g})= \tilde{g}^{4/5}/(2\pi)^{2/5}$.

\subsection{M(atrix)-Theory in Six Dimensions}
\label{ssect:sixdim}

For M(atrix)[$\widetilde{T}^5$], there are 25 moduli, corresponding to
the dimensionless coupling constant, 14 parameters of volume one
metrics on the dual torus, and 10 theta angles. This is precisely the
same number of moduli as appears in the case of IIA[$T^4$]. We expect
that it would not be too hard to use  an explicit
parameterization of $SO(5,5)/SO(5)\times SO(5)$ to determine the precise
mapping between the SYM and string moduli, as we have done above in the
higher dimensional cases. We will not, however, attempt that here.

Let us now consider the central charges. These are
 composed of 5~electric fluxes and 10~magnetic fluxes
transforming in the vector~$\mathbf{5}$ and the antisymmetric
tensor~$\mathbf{10}_a$ of $SL(5,\BZ)$. The spinor~$\mathbf{16}$ of
$SO(5,5,\BZ)$ decomposes under the $SL(5,\BZ)$ subgroup as
$\mathbf{1}\oplus \mathbf{5}\oplus \mathbf{10}_a$. Therefore, as was
also realized in~\cite{Fischler:Shrinking}, an additional state is
required to complete the desired U-duality multiplet.  From the
M-Theory picture, this must correspond to a transverse 5-brane which
completely wraps the 5-torus.  We note that, although a charge is
missing here, we do, in fact, have all of the modular parameters for
this compactification. It is nevertheless very important to give
an explicit construction of these transverse 5-brane states in the
M(atrix) model, as they are crucial to preserving U-duality.

\subsection{Compactification to Lower Dimensions}
\label{ssect:lower}

So far, we have considered compactifications for which, in the M-Theory
picture, transverse 5-branes do not contribute moduli. Nevertheless, we
did see how they are required to complete U-duality multiplets.  We
can summarize the dimensions of these M(atrix) SYM moduli spaces in
the same manner that we did for the IIA string in~\eqref{eq:IIAmoduli}
\begin{equation} \label{eq:SYMmoduli}
\begin{split}
\dim {\cal M}_{\text{M(atrix)}[\widetilde{T}^d]} =
& ~1~\text{dimensionless coupling constant} \\
& + \left( \frac{d(d+1)}{2} -1 \right)~\text{moduli of metrics of unit
volume} \\
& + \frac{d(d-1)(d-2)}{3!}~\text{theta angles}.
\end{split}
\end{equation}
Substituting $d=n+1$, we see that these formul\ae\ agree.

However, in spacetime dimensions $D\leq5$, the IIA theory has the
additional moduli given in~\eqref{eq:addIIAmoduli}. Thus the first
discrepancy occurs in M(atrix)[$\widetilde T^6$], where we are missing
one modulus in~\eqref{eq:SYMmoduli}.
In the M-Theory language, one can dualize the 3-form gauge field to a 6-form,
and the modulus in question corresponds to a constant expectation value for
this 6-form on $T^6$.

The difficulty, of course, is that this is the gauge field that couples to the
5-brane, and we do not have an explicit construction of the transverse 5-brane
in M(atrix)-Theory. If we did, we would, perhaps, be able to understand this
modulus.

Nevertheless, if we  continue to make the \textit{ans\"atz} of including the
central charges associated to the wrapped transverse 5-branes, we obtain the
correct representation of the U-duality group on the central charges
 for
M(atrix)[$\widetilde{T}^6$]. The central charge corresponds to
6~electric fluxes, 15~magnetic fluxes, and 6~wrapped transverse
5-branes. Our target is the $\mathbf{27}$ of $E_{6(6)}$, which
decomposes as $(\mathbf{6},\mathbf{2})\oplus
(\mathbf{15}_a,\mathbf{1})$ under the maximal subgroup
$SL(6,\BZ)\times SL(2,\BZ)$. Our collection satisfies this structure
under the $SL(6,\BZ)$ group of torus deformations, but the $SL(2,\BZ)$
symmetry is not manifest. The $SL(2,\BZ)$ symmetry mixes the electric
fluxes with the 5-branes, so, given the absence of an explicit
construction of the transverse 5-brane, it is not surprising that we
do not see it. Nevertheless, the algebraic structure that we do see is
evidence enough that the correct $E_{6(6)}$ U-duality is present.

Finally, we can consider M(atrix)[$\widetilde{T}^7$].
In this case, including the effects of the transverse 5-brane is
\textit{not} enough to insure the proper construction of the U-duality
representation.  We have explicitly 7~electric fluxes and 21~magnetic
fluxes, which fall into the $\mathbf{7}$ and $\mathbf{21}$ of the manifest
$SL(7,\BZ)$. These are what one expects from the decomposition of the
$\mathbf{56}$ of the U-duality group~$E_{7(7)}$, as
$\mathbf{7} +\mathbf{21} + \widetilde{\mathbf{21}}
+ \widetilde{\mathbf{7}}$ under the $SL(7,\BZ)$ subgroup. Our
discussion above would motivate us to add to the above central charges
21~wrapped transverse 5-branes, in the $\widetilde{\mathbf{21}}$ of
$SL(7,\BZ)$. There are, however, 7~additional charges, which according
to our summary in Table~\ref{tab:charges}, correspond to Kaluza-Klein
monopoles~\cite{{Sorkin:KKmonopole},{Gross-Perry:KKmonopole}}
in the corresponding M[$T^7$] picture. That is, they correspond to
geometries in which the 11-dimensional spacetime is not globally
$\BR^4\times T^7$. Rather, one has a \textit{non-trivial} $T^7$-bundle
over the 2-sphere at
spatial infinity. Near the origin, the fibration structure goes bad, but the
total space is non-singular. A non-trivial $S^1$-bundle over
$S^2$ corresponding to the monopole of charge 1 is given by the Hopf
fibration $S^3\to S^2$.  Since we have seven $S^1$s, there are 7 monopole
charges, which form the $\mathbf{7}$ of $SL(7,\BZ)$.

It is, to say the least, unclear how this structure is to be incorporated
into the M(atrix) theory description. Perhaps
this failure is related to the logarithmic divergence and the loss of
one-loop finiteness of the $7+1$-dimensional SYM theory that we
discussed in section~\ref{ssect:sevendim}. Another possibility is
that the missing charges may be related to the 6-brane charge found
in~\cite{Banks:BranesMat}. If such charges did appear in the 7+1-dimensional
SYM, they would, indeed, transform as the $\mathbf{7}$
of $SL(7,\BZ)$. However, by taking one of the radii of the
$\widetilde{T}^7$ to be
very small, one can argue that they should contribute a single central
charge for the SYM theory on $\widetilde{T}^6\times \BR$. But this would be a
disaster, as M(atrix)[$\widetilde{T}^6$] already gave the correct central
charges to agree with IIA[$T^5$]. Adding one more charge would ruin the
correspondence between M(atrix)-Theory and IIA string theory in $D=5$.

\section{Conclusions}

Our analysis of the toroidal M(atrix) SYM theories yields an explicit
construction of their moduli spaces. The correct global structure of
the moduli spaces is evident and agrees with that of the moduli spaces
of IIA compactifications. Moreover, the central charge of the M(atrix)
SYM theory transforms in the manner necessary to recover the
U-dualities of the corresponding string theories. We emphasize that
U-duality is, in general, only a symmetry of the BPS charges and is
not a symmetry of the action.

It is of particular interest to us to consider how these constructions
may be extended to more complicated scenarios of compactification.  In
particular, it is interesting to understand how far the M(atrix) model
can be used to extract information of the strong coupling limit of
different string theories and to provide a full description of
M-Theory. Clearly, the limitations of the M(atrix) model become
apparent when one compactifies on a large number of dimensions, indeed
on $\widetilde{T}^4$. At this point, the M(atrix)-Theory is a
$4+1$-dimensional SYM theory, over which we have little control from a
perturbative point of view. A more complete understanding of the
dynamics, such as information from IR fixed
points~\cite{{Rozali:Uduality},{Banks:StringsMat},{Fischler:Shrinking}},
is crucial to understanding these theories.  However, results in this
direction~\cite{Rozali:Uduality} suggest that these theories flow to
tensionless string theories. It is still not possible to apply these
to yield concrete statements which would extend our results.

In addition, on higher dimensional tori, the effect of M-Theory transverse
5-branes can no longer be ignored. Recent attempts to understand these
objects~\cite{{Lifschytz:Transverse},{Berkooz-Rozali:FiveBrane}} do not
appear to lend
themselves toward a description of the corresponding SYM moduli. The
treatment of Lifschytz~\cite{Lifschytz:Transverse} relied on repeated
T-dualization of the membrane. Since the full U-duality group is never
a symmetry of the SYM action, we cannot make use of this technique in
the approach we have outlined in this paper.  In the construction of
Berkooz and Rozali~\cite{Berkooz-Rozali:FiveBrane}, the 5-brane, wrapped
around the 5-torus, appears as a winding mode of a scalar field around
the 6th dimension of the 5+1-dimensional theory found
in~\cite{Rozali:Uduality}. This 5+1-dimensional theory is
essentially the same as that on the world-volume of the 5-brane and
involves self-dual tensors, and is not completely understood.  The theory
is potentially
anomalous~\cite{{Witten:FluxQuantization},{Witten:FiveBrane}} and the
self-duality of the tensor appears to spoil the possibility of
extending our approach within the SYM theories. At the time this
article is being written, these questions are unanswered.

The current absence of either a simple description of the dynamics or
an explicit construction of the transverse 5-brane in the M(atrix) SYM
formulation is certainly an impediment to obtaining a rich
understanding of M(atrix)-Theory. Indeed, recent work of Hashimoto and
Taylor~\cite{Hashimoto:Fluctuation} suggests that the SYM formulation
is too simple to describe certain configurations of tilted
branes~\cite{Polchinski:TASIDbrane} and branes intersecting at
angles~\cite{Berkooz:Angles}. In these cases, they found that one must
resort to a non-Abelian generalization of the Born-Infeld
action, such as that proposed by Tseytlin~\cite{Tseytlin:NonAbelian}.
Furthermore, it is even less evident how one might incorporate the presence of
Kaluza-Klein monopoles into the M(atrix) SYM, as will certainly be
necessary in $D=4$. What impact these results might have on
M(atrix)-Theory is nevertheless an exciting area of future
investigation.

\section*{Acknowledgements}

D.~B.~and R.~C.~would like to thank Willy Fischler for originally
suggesting that we study these toroidal moduli spaces and we thank him
for discussions about these and related matters. We also thank Moshe
Rozali for sharing the results of~\cite{Berkooz-Rozali:FiveBrane} with us
prior to e-printing and for many insightful discussions. We
additionally thank Vadim Kaplunovsky for discussions.

\renewcommand{\baselinestretch}{1.0} \normalsize


\bibliography{strings,m-theory,susy}
\bibliographystyle{utphys}

\end{document}